# Effective spin-mixing conductance of heavy-metal/ferromagnet interfaces

Lijun Zhu,[1*] Daniel C. Ralph,[1,2] and Robert A. Buhrman[1]
1. Cornell University, Ithaca, NY 14850
2. Kavli Institute at Cornell, Ithaca, New York 14853, USA

*Email: lz442@cornell.edu

The effective spin-mixing conductance ($G_{\text{eff}}^{\uparrow\downarrow}$) of a heavy metal/ferromagnet (HM/FM) interface characterizes the efficiency of the interfacial spin transport. Accurately determining $G_{\text{eff}}^{\uparrow\downarrow}$ is critical to the quantitative understanding of measurements of direct and inverse spin Hall effects. $G_{\text{eff}}^{\uparrow\downarrow}$ is typically ascertained from the inverse dependence of magnetic damping on the FM thickness under the assumption that spin pumping is the dominant mechanism affecting this dependence. Here we report that, this assumption fails badly in many in-plane magnetized prototypical HM/FM systems in the nm-scale thickness regime. Instead, the majority of the damping is from two-magnon scattering at the FM interface, while spin-memory-loss scattering at the interface can also be significant. If these two effects are neglected, the results will be an unphysical "giant" *apparent* $G_{\text{eff}}^{\uparrow\downarrow}$ and hence considerable underestimation of both the spin Hall ratio and the spin Hall conductivity in inverse/direct spin Hall experiments.
**Key words**: Spin-mixing conductance, spin-orbit coupling, spin memory loss, two-magnon scattering, spin Hall effect

Interfacial spin transport is at the root of many spintronic phenomena, e.g. spin-orbit torques (SOTs)[1,2], spin magnetoresistance (SMR)[3,4], the spin Seebeck effect (SSE)[5-7], and spin pumping [8-16] in heavy-metal /ferromagnet (HM/FM) systems. The key factor determining the spin transmission and spin backflow (SBF) of a HM/FM interface [17,18] is the effective spin-mixing conductance [19]

$$G_{\text{eff}}^{\uparrow\downarrow} = G_{\text{HM/FM}}^{\uparrow\downarrow}/(1+2G_{\text{HM/FM}}^{\uparrow\downarrow}/G_{\text{HM}}) \quad (1)$$

where $G_{\text{HM/FM}}^{\uparrow\downarrow}$ is the bare interfacial spin-mixing conductance, $G_{\text{HM}} = 1/\lambda_s \rho_{xx}$, $\rho_{xx}$, and $\lambda_s$ are the spin conductance, the resistivity, and the spin diffusion length of the HM layer, respectively. In inverse spin Hall effect (ISHE) experiments where spin currents are generated by spin pumping or the SEE, the measured voltage signals are proportional to $G_{\text{eff}}^{\uparrow\downarrow} \theta_{\text{SH}}$ [6-16]. Here $\theta_{\text{SH}}$ is the spin Hall ratio of the HM. For SOT experiments [20-23], the drift-diffusion analysis [4,17] predicts an interfacial spin transparency [19]

$$T_{\text{int}} = 2G_{\text{eff}}^{\uparrow\downarrow}/G_{\text{HM}} \leq 1 \quad (2)$$

when the HM thickness $d \gg \lambda_s$ and the interfacial spin-orbit coupling (ISOC) is negligible. Therefore, the measured dampinglike SOT efficiency per unit applied electric field is $\xi_{\text{DL}}^E \approx 2G_{\text{eff}}^{\uparrow\downarrow} \theta_{\text{SH}}/\rho_{xx}G_{\text{HM}}$. For SMR experiments, the measured resistance signals are proportional to $G_{\text{eff}}^{\uparrow\downarrow} \theta_{\text{SH}}^2$ [3,4]. As a consequence, for all of these techniques any errors in the determination of $G_{\text{eff}}^{\uparrow\downarrow}$ will directly result in incorrect evaluation of $\theta_{\text{SH}}$ and the spin Hall conductivity ($\sigma_{\text{SH}} \equiv (\hbar/2e)\theta_{\text{SH}}/\rho_{xx}$) of the HM; in general if $G_{\text{eff}}^{\uparrow\downarrow}$ is overestimated, $\theta_{\text{SH}}$ and $\sigma_{\text{SH}}$ will be underestimated.

In practice, $G_{\text{eff}}^{\uparrow\downarrow}$, or equivalently $g_{\text{eff}}^{\uparrow\downarrow} = G_{\text{eff}}^{\uparrow\downarrow} h/e^2$, for a HM/FM system is typically determined by measuring the FM thickness ($t_{\text{FM}}$) dependence of the damping ($\alpha$) of in-plane magnetized bilayers based on the standard model where the $t_{\text{FM}}$ dependence is attributed only to the enhancement of $\alpha$ by spin pumping into the HM layer [7-12,24-28], i.e.

$$\alpha = \alpha_{\text{int}} + G_{\text{eff},\alpha}^{\uparrow\downarrow} \frac{g\mu_B h}{4\pi M_s e^2} t_{\text{FM}}^{-1} \quad (3)$$

where $\alpha_{\text{int}}$ is the thickness-independent "intrinsic" damping of the FM layers, $M_s$ the saturation magnetization of the FM layers, $g$ the $g$-factor, $\mu_B$ the Bohr magnetron, and $h$ Planck's constant, respectively. The apparent values of $G_{\text{eff},\alpha}^{\uparrow\downarrow}$ obtained by this method have been widely used to estimate $\theta_{\text{SH}}$ and $\sigma_{\text{SH}}$ in many spin pumping/ISHE, SEE/ISHE, SMR and SOT experiments [4,6,8-11,13-16,24].

In this letter, we report that spin pumping is a relatively minor contribution to $\alpha$ for the most commonly-studied in-plane magnetized HM/FM systems in the nm thickness and GHz frequency regions that are of most interest for spintronics. In contrast, two-magnon scattering (TMS)[29,30] predominantly determines the $t_{\text{FM}}$ dependence of $\alpha$. When ISOC is sufficiently strong, the second largest contribution can be spin memory loss (SML)[12,31-35]. Neglecting TMS and SML, particularly the former, when analyzing measurements of $\alpha$ in HM/FM systems gives unphysical "giant" estimates for $G_{\text{eff}}^{\uparrow\downarrow}$ and hence incorrect quantification of spin-dependent transport phenomena across HM/FM interfaces and large errors in the determination of $\theta_{\text{SH}}$ and $\sigma_{\text{SH}}$.

For this study we use six different series of sputter-deposited in-plane magnetized Pt/FM samples as examples (see Table 1): (1) as-grown Pt/NiFe ($Ni_{81}Fe_{19}$); (2) as-grown Pt/FeCoB ($Fe_{60}Co_{20}B_{20}$); (3) as-grown Pt/Co; (4) Pt/Co, annealed at 300 ºC; (5) Pt/Co, annealed at 350 ºC; and (6) Pt/Co, annealed at 450 ºC. The Pt thickness is 4 nm in all cases, while in each series $t_{\text{FM}}$ was varied over a sufficient range to reveal the damping behavior. $\alpha$ was determined by spin-torque ferromagnetic resonance [36]. See Supplementary Materials [37] for more information on the samples and experimental methods.

Table 1. Details of the Pt/FM sample series. $T_a$ is the annealing temperature. The estimates for $\alpha_{\text{int}}$ labeled "No TMS" are obtained from the intercepts of linear fits of $\alpha$ to $t_{\text{FM}}^{-1}$ (Eq. 3) that neglect TMS (Figs. 1(a) and 1(b)). The estimates labeled "With TMS" are the results of full fits to Eq. (4) taking into account the TMS contribution (Figs. 2(b)-2(d)).

| Series | FM | $T_a$ (ºC) | $K_s$ (erg/cm$^2$) | $\alpha_{\text{int}}$ | |
|---|---|---|---|---|---|
| | | | | No TMS | With TMS |
| 1 | NiFe | NA | 0.31±0.09 | 0.006 | 0.011 |
| 2 | FeCoB | NA | 0.84±0.06 | 0.003 | 0.006 |
| 3 | Co | NA | 0.90±0.07 | 0.001 | 0.011±0.001 |
| 4 | Co | 300 | 1.62±0.02 | 0.003±0.003 | 0.010 ±0.002 |
| 5 | Co | 350 | 2.52±0.02 | -0.014±0.007 | 0.008 ±0.002 |
| 6 | Co | 450 | 3.27±0.02 | -0.030±0.011 | 0.010 ±0.006 |

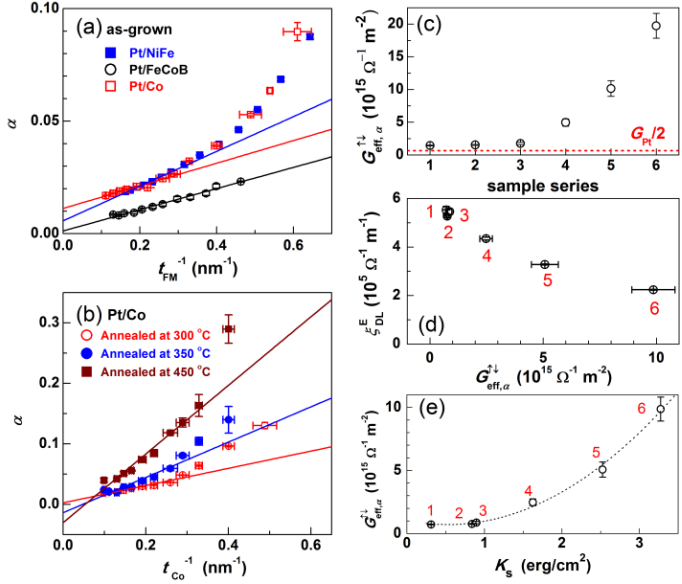

Fig. 1. Dependence of $\alpha$ on $t_{FM}^{-1}$ for (a) the as-grown Pt/NiFe, Pt/FeCoB, and Pt/Co sample series and (b) the annealed Pt/Co sample series. The straight lines represent linear fits in the thick FM region (Eq. (3)); (c) $g_{eff,\alpha}^{\uparrow\downarrow}$ and $G_{eff,\alpha}^{\uparrow\downarrow}$ for the different Pt/FM interfaces determined from the linear fits of $\alpha$ vs $t_{FM}^{-1}$ (Eq. (3)). The red dashed line represents $G_{Pt}/2 = 0.65 \times 10^{15}$ $\Omega^{-1}$ $m^{-2}$, an *upper* bound for the true value of $G_{eff}^{\uparrow\downarrow}$. (d) $\xi_{DL}^E$ vs $G_{eff,\alpha}^{\uparrow\downarrow}$. (e) $G_{eff,\alpha}^{\uparrow\downarrow}$ vs $K_s$. The numbers 1-6 in (d) and (e) label the sample series.

In Figs. 1(a) and 1(b) we plot $\alpha$ as a function of $t_{FM}^{-1}$ for the as-grown and annealed sample series, respectively. While $\alpha$ for the Pt/FeCoB samples varies quasi-linearly with $t_{FM}^{-1}$ over the full $t_{FM}$ range, $\alpha$ for Pt/NiFe and Pt/Co sample series deviates markedly from the linear $t_{FM}^{-1}$ scaling when $t_{FM}$ is small. Focusing first on the large $t_{FM}$ regime where $\alpha$ can be fit phenomenologically by a linear $t_{FM}^{-1}$ dependence (*i.e.*, $t_{FM}$> 3.5 nm for Pt/NiFe and Pt/Co series; $t_{FM}$>2.2 nm for Pt/FeCoB series), we determined $G_{eff,\alpha}^{\uparrow\downarrow}$ (Fig. 1(c)) and $g_{eff,\alpha}^{\uparrow\downarrow}$ for the different sample series from the fits of $\alpha$ to Eq. (3). $G_{eff,\alpha}^{\uparrow\downarrow}$ increases from $1.5 \times 10^{15}$ $\Omega^{-1}$ $m^{-2}$ for the Pt/NiFe and Pt/FeCoB samples (series 1 and 2), to $1.8 \times 10^{16}$ $\Omega^{-1}$ $m^{-2}$ for as-grown Pt/Co samples (series 3), and to $2.0 \times 10^{16}$ $\Omega^{-1}$ $m^{-2}$ for Pt/Co annealed at 450 °C (sample series 6). Corresponding values of $g_{eff,\alpha}^{\uparrow\downarrow}$ are $4.0 \times 10^{19}$ $m^{-2}$ (series 1 and 2), $4.5 \times 10^{19}$ $m^{-2}$ (series 3), and $5.1 \times 10^{20}$ $m^{-2}$ (series 6). These $G_{eff,\alpha}^{\uparrow\downarrow}$ ($g_{eff,\alpha}^{\uparrow\downarrow}$) values for our as-grown Pt/FM samples are comparable to those reported from damping measurements on similar structures in the literature [9,11,13,16,24,38]. However, all of these values of $G_{eff,\alpha}^{\uparrow\downarrow}$ are markedly larger than the expected value $G_{eff}^{\uparrow\downarrow}=0.31 \times 10^{15}$ $\Omega^{-1}$ $m^{-2}$ as calculated with Eq. (1) using the *ab-initio* prediction $G_{Pt/Co}^{\uparrow\downarrow}= 0.59 \times 10^{15}$ $\Omega^{-1}$ $m^{-2}$ [18] and the experimentally determined $G_{Pt} = 1.3 \times 10^{15}$ $\Omega^{-1}$ $m^{-2}$ [20].

Moreover, while Eq. (1) requires that $G_{eff}^{\uparrow\downarrow} \leq G_{Pt}/2$, all the values of $G_{eff,\alpha}^{\uparrow\downarrow}$ we obtain using Eq. (3) are substantially larger than $G_{Pt}/2$, with the ratio $G_{eff,\alpha}^{\uparrow\downarrow}/(G_{Pt}/2)$ as large as 30 for Pt/Co series annealed at 450 °C! When $G_{eff}^{\uparrow\downarrow} > G_{HM}/2$, the value of $G_{HM/FM}^{\uparrow\downarrow} = G_{eff}^{\uparrow\downarrow}/(1-2\, G_{eff}^{\uparrow\downarrow}/G_{HM})$ will be negative, which is unphysical. Our group has also observed values of $G_{eff}^{\uparrow\downarrow}$ much larger than $G_{HM}/2$ from damping studies of Pt/CoFe [32] and PtMn/(FeCoB,Co) systems [28]. A giant $G_{eff}^{\uparrow\downarrow}$ may only be consistent with the bound ($G_{eff}^{\uparrow\downarrow} < G_{HM}/2$) in Eq. (1) if $\lambda_s$ is much shorter (e.g., < 0.06 nm for Pt/Co annealed at 450 °C) than determined by independent measurements (~2 nm)[11]. However, this is also unphysical because it implies a value for $G_{HM/FM}^{\uparrow\downarrow}$ that is much greater than the Sharvin conductance of Pt ($G_{Sh}=0.68 \times 10^{15}$ $\Omega^{-1}$ $m^{-2}$)[39]. Note that the drift-diffusion model [4,17] and Eq. (1) require $G_{eff}^{\uparrow\downarrow} < G_{HM/FM}^{\uparrow\downarrow} \leq G_{Sh}$. Notably, the failure of the assumption of spin pumping dominating the FM thickness dependence of $\alpha$ and the deduced unphysical, giant $G_{eff,\alpha}^{\uparrow\downarrow}$ are not particular to Pt/FM systems, but is generally observed for other HM/FM systems, including Pd$_{0.25}$Pt$_{0.75}$/(Co,FeCoB), Au$_{0.25}$Pt$_{0.75}$/(Co,FeCoB), Pd/Co, and W/FeCoB [37].

Variations of $\xi_{DL}^E$ between the different Pt/FM series provide another illustration of the danger of misinterpreting these giant $G_{eff,\alpha}^{\uparrow\downarrow}$. Figure 1(d) plots $\xi_{DL}^E$ from harmonic response measurements [19,20,40] on the representative samples of each series (i.e., Pt 4/NiFe 1.8 for series 1, Pt 4/FeCoB 2.8 for series 2, Pt 4/Co 3.2 for series 3-6) as a function of $G_{eff,\alpha}^{\uparrow\downarrow}$. For an ideal Pt/FM interface where $T_{int}$ is set only by SBF, according to Eq. (2) a large $G_{eff}^{\uparrow\downarrow}$ should favor a high $\xi_{DL}^E$. However, we find experimentally that $\xi_{DL}^E$ decreases substantially and monotonically with increasing $G_{eff,\alpha}^{\uparrow\downarrow}$. Both the unphysically-large $G_{eff,\alpha}^{\uparrow\downarrow}/(G_{Pt}/2)$ ratios and the anti-correlation between $\xi_{DL}^E$ and $G_{eff,\alpha}^{\uparrow\downarrow}$ provide unambiguous evidence that the values of $G_{eff,\alpha}^{\uparrow\downarrow}$ determined from the standard spin-pumping model, Eq. (3), are not accurate estimates of $G_{eff}^{\uparrow\downarrow}$ defined by Eq. (1).

It has been established that SML due to the ISOC at the Pt/FM interfaces [12,31-35] provides an additional spin sink that can increase $\alpha$ and degrade $T_{int}$ for the HM/FM interface. However, we find that the giant $G_{eff,\alpha}^{\uparrow\downarrow}$ values are not due primarily to SML. First, spin pumping into the SML interface should yield a $t_{FM}^{-1}$ dependence for $\alpha$ and cannot explain the strong deviation from a $t_{FM}^{-1}$ dependence that we find in the thin $t_{FM}$ regime (Figs. 1(a) and 1(b)). Second, if the enhancement in $G_{eff,\alpha}^{\uparrow\downarrow}$ by a factor of 20 from sample series 1 to 6 (Fig. 1(d)) were due to SML, $\xi_{DL}^E$ should be reduced by a similarly-large factor, rather than being reduced only by 25%. Finally, we find that $G_{eff,\alpha}^{\uparrow\downarrow}$ scales approximately as the *square* of the interfacial magnetic anisotropy energy density ($K_s$) at the Pt/FM interfaces (Fig. 1(e)) as determined by measuring the effective demagnetization field $M_{eff}$ using FMR and fitting to a $t_{FM}^{-1}$ dependence [37]. In contrast, theory predicts that the contribution from SML should be linear in the ISOC strength [14] and therefore $K_s$ [33].

A third possible contribution to enhanced damping that has been seldom considered when analyzing interfacial spin transport is TMS [29] due to magnetic defects (roughness) at the interfaces (see Fig. S5 in [37]). As we next discuss, TMS dominates the enhancement of $\alpha$ in the Pt/FM heterostructures. A signature for the TMS contribution to the damping ($\alpha_{TMS}$) is that, to the first approximation, it is a parabolic function of the interfacial perpendicular magnetic anisotropy field $2K_s/M_s t_{FM}$ [29,30]. If $\alpha_{TMS}$ is significant, the total damping is given approximately by $\alpha = \alpha_{int} + \alpha_{SP} + \alpha_{TMS}$ or

$$\alpha = \alpha_{\text{int}} + (G_{\text{eff}}^{\uparrow\downarrow} + G_{\text{SML}}) \frac{g\mu_B h}{4\pi M_s e^2} t_{\text{FM}}^{-1} + \beta_{\text{TMS}} t_{\text{FM}}^{-2} \quad (4)$$

where the second term is the combined contribution from spin pumping into the Pt layer and from SML at the interface, which for convenience we parameterize as an "effective SML conductance" $G_{\text{SML}}$[16,27], and $\beta_{TMS}$ is a coefficient that depends on both $(K_s/M_s)^2$ and the density of magnetic defects at the FM surfaces [29,30](We provide further justification for the $t_{\text{FM}}^{-2}$ dependence of the TMS term in [37]).

To properly fit the damping data to Eq. (4) and disentangle the different contributions, we must first estimate $G_{\text{eff}}^{\uparrow\downarrow}$ for the spin pumping into the Pt layer and $G_{\text{SML}}$ for spin pumping into the SML interface for the different Pt/FM series. We note that the expected value of $G_{\text{eff}}^{\uparrow\downarrow} = 0.31\times10^{15}$ $\Omega^{-1}$ m$^{-2}$ for Pt/Co interface (Eq. (1)) is in reasonable agreement with the experimental value of $G_{\text{eff},\alpha}^{\uparrow\downarrow}$ obtained in Pt/FM samples [32,38] where the interfaces were engineered to reduce ISOC, and thereby minimize SML and TMS. For example, the damping measurements (Eq. (3)) have yielded $G_{\text{eff},\alpha}^{\uparrow\downarrow} = 0.38\times10^{15}$ $\Omega^{-1}$ m$^{-2}$ for unannealed CoFe/Pt bilayers [32] and $G_{\text{eff},\alpha}^{\uparrow\downarrow} = 0.25\times10^{15}$ $\Omega^{-1}$ m$^{-2}$ for a Pt/FeCoB bilayers where ISOC (TMS and SML) was diminished by inserting a 0.5 nm Hf passivation spacer [37]. This indicates that $G_{\text{eff}}^{\uparrow\downarrow} = 0.31\times10^{15}$ $\Omega^{-1}$ m$^{-2}$ is a reasonable approximation for all the Pt/FM samples. To account for the SML contribution, we identify the reduction in $\xi_{\text{DL}}^E$ with increasing ISOC (Fig. 1(d), Fig. 2(a)) as being due to SML, and assume that the fraction of spin current absorbed by SML at the interface is the same for the spin-pumping (FM→HM) and SOT (HM→FM) processes. Based on this approximation, we obtain

$$G_{\text{SML}} \approx G_{\text{eff}}^{\uparrow\downarrow} \frac{\xi_{\text{DL}}^E(\text{no SML}) - \xi_{\text{DL}}^E}{\xi_{\text{DL}}^E}. \quad (5)$$

Previous work [33] has established that $\xi_{\text{DL}}^E$ deceases linearly with ISOC strength at the HM/FM interfaces ($K_s^{\text{HM/FM}}$) due to SML. In Fig. 2(a), we determine the baseline value of $\xi_{\text{DL}}^E$(no SML) = $(5.9\pm0.1) \times10^5$ $\Omega^{-1}$ m$^{-1}$ as determined from the linear $K_s^{\text{Pt/Co}}$ dependence of $\xi_{\text{DL}}^E$ for Pt/Co bilayers. Using the measured $\xi_{\text{DL}}^E$, Eq. (5) yields the values of $G_{\text{SML}}$ shown in Fig. 2(b). We find that $G_{\text{SML}}$ is negligible relative to $G_{\text{eff}}^{\uparrow\downarrow}$ for the unannealed samples, but as a function of increasing annealing temperature $G_{\text{SML}}$ becomes comparable to and then larger than $G_{\text{eff}}^{\uparrow\downarrow}$.

With the values of $G_{\text{eff}}^{\uparrow\downarrow}$ and $G_{\text{SML}}$ in Fig. 2(b), the damping data for all the Pt/FM series can be fit well by Eq. (4) over the whole range of $t_{\text{FM}}$ studied, using the two fitting parameters $\beta_{\text{TMS}}$ and $\alpha_{\text{int}}$ (see Figs. 2(c)-2(e)). As shown in Figs. 2(c) and 2(d), $\alpha_{\text{TMS}}$ (red line) for both the Pt/NiFe and Pt/FeCoB sample series is larger than $\alpha_{\text{SP}}$ (blue line) in the whole range of $t_{\text{FM}}$. For the Pt/Co samples (series 3-6), the dominant parabolic scaling of $\alpha$ with $t_{\text{FM}}^{-1}$ becomes increasingly stronger with increasing annealing temperature (and therefore increasing $K_s$) (Fig. 2(e)). These observations demonstrate that TMS constitutes the largest thickness-dependent contribution to $\alpha$ for all of the Pt/FM systems we have examined, even for Pt/FeCoB ($K_s = 0.84\pm0.06$ erg/cm$^2$) where $\alpha$ appears to vary quasi-linearly with $t_{\text{FM}}^{-1}$ (Fig. 1(a)). We also find that, for Pt/Co samples (series 3-6) with the similar structural roughness at the interfaces, the TMS coefficient $\beta_{TMS}$ determined from the best fits in Figs. 2(c)-2(e) scales monotonically with $(2K_s/M_s)^2$ (Fig.

2(f)), in good agreement with the TMS mechanism [29,30]. $\beta_{TMS}$ for Pt/FeCoB samples is~3 time smaller than that of the as-grown Pt/Co and Pt/NiFe samples despite of their similar values of $(2K_s/M_s)^2$, which indicates a smaller magnetic roughness at the amorphous FeCoB surfaces than at the polycrystalline NiFe and Co surfaces, where the latter two show columnar growth on top of Pt (see Fig. S5). Because $\alpha_{\text{SP}} \ll \alpha_{\text{TMS}}$ for most FM thicknesses (particularly in the small $t_{\text{FM}}$ range), our conclusions are not sensitive to the details of the fitting procedure. As shown in Fig. 2(f), the fits of $\alpha$ to Eq. (4) give essentially the same $\beta_{TMS}$ values for the different Pt/FM series whether we assume $G_{\text{Pt/FM}}^{\uparrow\downarrow} = 0$, $0.59\times10^{15}$ (theory [17]), or $1.18\times10^{15}$ $\Omega^{-1}$ m$^{-2}$. Since $\alpha_{\text{TMS}}$ dominates $\alpha$, the accuracy of the above conclusions are robust against any potential limitations of the drift-diffusion analysis [17,18].

For the HM/YIG (e.g., HM=Pt, Ta, W, Cu) bilayers [41], we find both conditions of Eq. (1) (i.e., $G_{\text{eff}}^{\uparrow\downarrow} < G_{\text{HM}}/2$ and $G_{\text{eff}}^{\uparrow\downarrow} < G_{\text{HM/FM}}^{\uparrow\downarrow} \leq G_{\text{Sh}}$) may be satisfied only when the real $G_{\text{HM}}$ values are much smaller than used in literature [41](see [37]). This most likely suggests that the TMS is weak in those HM/YIG bilayers where the YIG layers are very thick (>20 nm [41]), but that the $\lambda_s$ values of the HMs were considerably overestimated and thus the $\theta_{SH}$ values were underestimated in those spin pumping/ISHE experiments.

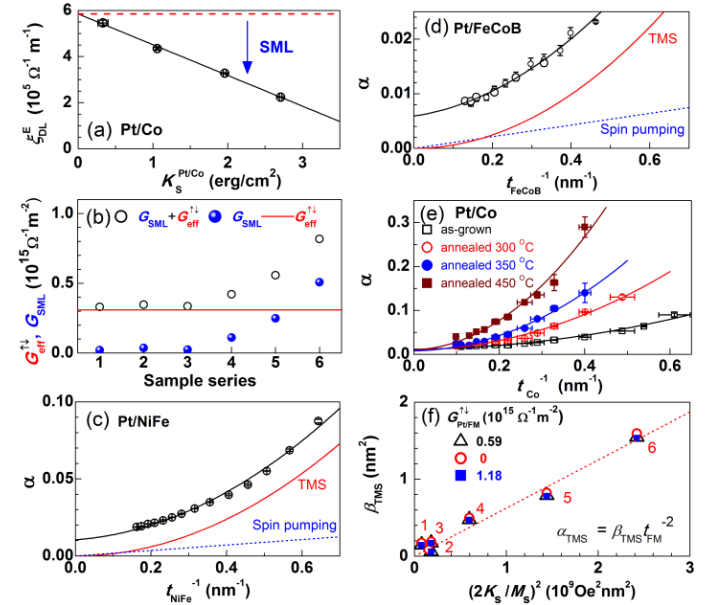

Fig. 2. (a) Reduction of $\xi_{\text{DL}}^E$ for Pt 4/Co 3.2 bilayers with increasing $K_s^{\text{Pt/Co}}$, indicating an extrapolated value of $\xi_{\text{DL}}^E = 5.9\times10^5$ $\Omega^{-1}$ m$^{-1}$ for zero SML (and zero ISOC) at the Pt/Co interface. (b) $G_{\text{SML}}$, $G_{\text{eff}}^{\uparrow\downarrow}$, and the sum of the two. Damping for (c) Pt/NiFe, (d) Pt/FeCoB, and (e) Pt/Co plotted as a function of $t_{\text{FM}}^{-1}$. In (c) and (d), the black lines represent best fits of the data to Eq. (4) including the TMS contribution; the solid red and the dashed blue lines indicate separately the TMS contribution and the total contribution of spin pumping into the Pt and the SML layer. In (e), the solid lines represent best fits of the data to Eq. (4). The intercepts in (c)-(e) indicate the intrinsic damping of the FM layer (the values in Table 1 labeled "With TMS"). (f) $\beta_{TMS}$ vs $(2K_s/M_s)^2$ as determined by fits to Eq. (4) assuming different values of $G_{\text{Pt/FM}}^{\uparrow\downarrow}$: 0, $0.59\times10^{15}$, and $1.18\times10^{15}$ $\Omega^{-1}$ m$^{-2}$. The red numbers 1-6 in (f) label the sample series.

In a damping analysis for HM/FM bilayers where only spin pumping is considered, the $t_{FM}$-independent contribution, the intercept of the linear $t_{FM}^{-1}$ fit of $\alpha$ using Eq. (3), is typically ascribed to the "intrinsic" $\alpha_{int}$ of the FM layer. However, the *apparent* $\alpha_{int}$ obtained from such linear $t_{FM}^{-1}$ fits of $\alpha$ is often unphysically small or even negative, e.g., being negative for Pt/Co and Pt/NiFe in Ref. [32]. As revealed by our numerical simulation [37], this is because $\alpha_{TMS}$ always contributes to a negative intercept in the fit of $\alpha$ to Eq. (3) in the thick FM region. As we summarize in Table 1, for our samples we observe similar fitting behaviors, the linear $t_{FM}^{-1}$ fits of total damping $\alpha$ in the thick FM region yield very small $\alpha_{int}$ for the as-grown Pt/FM sample sets (Fig. 1(a)) and negative $\alpha_{int}$ for annealed Pt/Co sample sets (Fig. S4). Including the effects of TMS on $\alpha$ (Figs. 2(c)-2(e)) resolves this problem and yields reasonable $\alpha_{int}$ values (~0.011 for NiFe, ~0.006 for FeCoB, and ~0.010 for Co), which are in accord with the literature values for free-standing thin films of these materials [42,43]. We observe similar effects in many other HM/FM sample series (see [37] for more examples on ($Pd_{0.25}Pt_{0.75}$, $Au_{0.25}Pt_{0.75}$, Pd)/Co and ($Pd_{0.25}Pt_{0.75}$, $Au_{0.25}Pt_{0.75}$, W)/FeCoB sample series). This finding therefore indicates that TMS must be taken into account when estimating $\alpha_{int}$ of a FM from its thickness dependence.

The understanding that our analysis provides about the relative strength of TMS, SML, and spin-pumping, and how these processes depend on the ISOC strength, has wide-ranging implications for correctly understanding spin current transport at HM/FM interfaces. As aforementioned, $\xi_{DL}^{E}$ can vary inversely with $G_{eff,\alpha}^{\uparrow\downarrow}$ even though according to Eq. (2) it should increase with increases in true $G_{eff}^{\uparrow\downarrow}$. Similarly, voltage signals in spin-pumping/ISHE measurements in Pt/FM samples can vary inversely with $G_{eff,\alpha}^{\uparrow\downarrow}$ [44] even though they are expected to scale $\propto G_{eff}^{\uparrow\downarrow}\theta_{SH}$. These puzzles are resolved if the dominant contribution to $G_{eff,\alpha}^{\uparrow\downarrow}$ is TMS, which does not affect spin transport across the interface. In addition, previous observations of an increase of $G_{eff,\alpha}^{\uparrow\downarrow}$ with the FM roughness [45], and scaling of $G_{eff,\alpha}^{\uparrow\downarrow}$ at HM/CoFeB interfaces with the interfacial Dzyaloshinskii-Moriya interaction constant (a factor that is proportional to the ISOC strength)[46] can be natural consequences of TMS. We note that $\alpha$ has been reported to be much larger in some HM/FM structures (e.g., Pt/CoFe [19] or PtMn/FeCoB [28]) than in the corresponding reversed order structures (i.e., FM/HM) despite their similar SOT strengths. This is consistent with a stronger $\alpha_{TMS}$ due to a larger magnetic roughness when the FM is grown on top of the HM. The dominant role of TMS in determining $\alpha$ in the thin HM/FM systems also explains observations that the reduction of ISOC at HM/FM interface by a Hf atomic layer insertion can dramatically reduce $\alpha$ in Pt/Co, Pt/FeCoB, and W/FeCoB systems without materially decreasing $T_{int}$ for the diffusion of spins from the HM layer into the FM layer [33,37,47].

Our results also indicate an critical need to re-evaluate all measurements of $\theta_{SH}$ and $\sigma_{SH}$ that utilize $G_{eff,\alpha}^{\uparrow\downarrow}$ from damping measurements to estimate the true $G_{eff}^{\uparrow\downarrow}$. Given that the measured signal strengths scale as $\propto G_{eff}^{\uparrow\downarrow}\theta_{SH}$ in spin-pumping/ISHE and SSE/ISHE measurements and $\propto G_{eff}^{\uparrow\downarrow}\theta_{SH}^{2}$ in SMR measurements, the common use of $G_{eff,\alpha}^{\uparrow\downarrow}$ as a proxy for $G_{eff}^{\uparrow\downarrow}$ means that most values of $\theta_{SH}$ in the literature determined by these techniques significantly underestimate the correct values. SOT measurements are often quoted as providing only a lower bound on $\theta_{SH}$ or $\sigma_{SH}$, assuming only that $T_{int} \leq 1$. While these lower bounds remain accurate, the improved understanding of $G_{eff}^{\uparrow\downarrow}$ from our analysis allows a more-confident quantification of $T_{int}$ so as to provide accurate measurements of $\theta_{SH}$ or $\sigma_{SH}$ using SOT experiments. By the analysis associated with Fig. 2, the values of $T_{int}$ for our Pt 4/FM sample series vary with the strength of ISOC such that $T_{int} \approx 0.38$ for samples series 1-3, 0.30 for series 4, 0.23 for series 5, and 0.16 for series 6. In all cases, our data are consistent with $\approx 1.5\times 10^{6}$ ($\hbar/2e$) $\Omega^{-1}$ $m^{-1}$ or $\theta_{SH} = 0.64$ within Pt given an average resistivity of $\rho_{xx} = 40$ $\mu\Omega$ cm. These values are consistent with previous estimates in [33].

In conclusion, two-magnon scattering rather than spin pumping is the dominant contribution to the FM-thickness dependence of $\alpha$ for in-plane-magnetized HM/FM systems in the nm thickness and GHz frequency regions important for spintronics. SML at the interface can also play an important role in affecting both $\alpha$ and $T_{int}$ when ISOC is strong. Neglecting the influence of TMS and SML, particularly the former, can lead to unphysical giant estimates for $G_{eff}^{\uparrow\downarrow}$. A correct calculation of $\theta_{SH}$ and $\sigma_{SH}$ therefore requires careful determination of the strength of both TMS and SML. Our findings also indicate that ISOC and magnetic roughness should be minimized in technological applications that benefit from low $\alpha$.

This work was supported in part by the Office of Naval Research (N00014-15-1-2449) and by the NSF MRSEC program (DMR-1719875) through the Cornell Center for Materials Research. This work was performed in part at the Cornell NanoScale Facility, an NNCI member supported by NSF Grant ECCS-1542081.

**References**
[1] L. Liu, C.-F. Pai, Y. Li, H. W. Tseng, D. C. Ralph, and R. A. Buhrman, Spin-torque switching with the giant Spin hall effect of tantalum, Science **336**, 555 (2012).
[2] V. E. Demidov, S. Urazhdin, H. Ulrichs, V. Tiberkevich, A. Slavin, D. Baither, G. Schmitz and S. O. Demokritov, Magnetic nano-oscillator driven by pure spin current, Nat. Mater. 11, 1028–1031 (2012).
[3] C. O. Avci, K. Garello, A. Ghosh, M. Gabureac, S. F. Alvarado, and P. Gambardella, Unidirectional spin Hall magnetoresistance in ferromagnet/normal metal bilayers, Nat. Phys. 11, 570–575 (2015).
[4] Y.-T. Chen, S. Takahashi, H. Nakayama, M. Althammer, S. T. B. Goennenwein, E. Saitoh, and G. E. W. Bauer, Theory of spin Hall magnetoresistance, Phys. Rev. B 87, 224401 (2013).
[5] K. Uchida, S. Takahashi, K. Harii, J. Ieda, W. Koshibae, K. Ando, S. Maekawa, and E. Saitoh, Observation of the spin Seebeck effect, Nature **455**, 778–781(2008).
[6] D. Meier, D. Reinhardt, M. van Straaten, C. Klewe, M. Althammer, M. Schreier, S. T. B. Goennenwein, A. Gupta, M. Schmid, C. H. Back, J.-M. Schmalhorst, T. Kuschel & G. Reiss, Longitudinal spin Seebeck effect contribution in transverse spin Seebeck effect experiments in Pt/YIG and Pt/NFO, Nat. Commun. 6, 8211 (2015).
[7] W. Lin, K. Chen, S. Zhang, C. L. Chien, Enhancement of Thermally Injected Spin Current through an Antiferromagnetic Insulator, Phys. Rev. Lett. 116, 186601 (2016)

[8] J. C. Rojas Sánchez, L. Vila, G. Desfonds, S. Gambarelli, J. P. Attané, J. M. De Teresa, C. Magén, A. Fert, Spin-to-charge conversion using Rashba coupling at the interface between non-magnetic materials, Nat. Commun. 4, 2944 (2013).

[9] X. Tao, Q. Liu, B. Miao, R. Yu, Z. Feng, L. Sun, B. You, J. Du, K. Chen. S. Zhang, L. Zhang, Z. Yuan, D. Wu, and H. Ding, Self-consistent determination of spin Hall angle and spin diffusion length in Pt and Pd: The role of the interface spin loss, Sci. Adv. 4, eaat1670 (2018).

[10] B. Heinrich, C. Burrowes, E. Montoya, B. Kardasz, E. Girt, Y.-Y. Song, Yiyan Sun, and M. Wu, Spin Pumping at the Magnetic Insulator (YIG)/Normal Metal (Au) Interfaces, Phys. Rev. Lett. 107, 066604 (2011).

[11] M. Obstbaum, M. Decker, A. K. Greitner, M. Haertinger, T. N. G. Meier, M. Kronseder, K. Chadova, S. Wimmer, D. Ködderitzsch, H. Ebert, and C. H. Back, Tuning Spin Hall Angles by Alloying, Phys. Rev. Lett. **117**, 167204 (2016).

[12] K. Chen and S. Zhang, Spin Pumping in the Presence of Spin-Orbit Coupling, Phys. Rev. Lett. 114, 126602 (2015).

[13] F. D. Czeschka, L. Dreher, M. S. Brandt, M. Weiler, M. Althammer, I.-M. Imort, G. Reiss, A. Thomas, W. Schoch, W. Limmer, H. Huebl, R. Gross, and S. T. B. Goennenwein, Scaling Behavior of the Spin Pumping Effect in Ferromagnet-Platinum Bilayers, Phys. Rev. Lett. 107, 046601 (2011).

[14] O. Mosendz, J. E. Pearson, F. Y. Fradin, G. E. W. Bauer, S. D. Bader, and A. Hoffmann, Quantifying Spin Hall Angles from Spin Pumping: Experiments and Theory, Phys. Rev. Lett. 104, 046601 (2010).

[15] F. D. Czeschka, L. Dreher, M. S. Brandt, M. Weiler, M. Althammer, I.-M. Imort, G. Reiss, A. Thomas, W. Schoch, W. Limmer, H. Huebl, R. Gross, and S. T. B. Goennenwein, Scaling Behavior of the Spin Pumping Effect in Ferromagnet-Platinum Bilayers, Phys. Rev. Lett. 107, 046601(2011).

[16] J.-C. Rojas-Sánchez, N. Reyren, P. Laczkowski, W. Savero, J.-P. Attané, C. Deranlot, M. Jamet, J.-M. George, L. Vila, and H. Jaffrès, Spin pumping and inverse spin Hall effect in platinum: the essential role of spin-memory loss at metallic interfaces, Phys. Rev. Lett. **112**, 106602 (2014).

[17] P. M. Haney, H. W. Lee, K. J. Lee, A. Manchon, and M. D. Stiles, Current induced torques and interfacial spin-orbit coupling: Semiclassical modeling, Phys. Rev. B **87**, 174411 (2013).

[18] V. P. Amin and M. D. Stiles, Spin transport at interfaces with spin-orbit coupling: Phenomenology, Phys. Rev. B **94**, 104420 (2016).

[19] C.-F. Pai, Y. Ou, L. H. Vilela-Leao, D. C. Ralph, R. A. Buhrman, Dependence of the efficiency of spin Hall torque on the transparency of Pt/ferromagnetic layer interfaces, Phys. Rev. B **92**, 064426 (2015).

[20] M.-H. Nguyen, D. C. Ralph, R. A. Buhrman, Spin torque study of the spin Hall conductivity and spin diffusion length in Platinum thin films with varying resistivity, Phys. Rev. Lett. **116**, 126601 (2016).

[21] L. Zhu, D. C. Ralph, and R. A. Buhrman, Highly efficient spin current generation by the spin Hall effect in $Au_{1-x}Pt_x$, Phys. Rev. Applied **10**, 031001 (2018).

[22] L. J. Zhu, K. Sobotkiewich, X. Ma, X. Li, D. C. Ralph, and R. A. Buhrman, Strong dampinglike spin-orbit torque and tunable Dzyaloshinskii-Moriya interaction generated by low-resistivity $Pd_{1-x}Pt_x$, Adv. Fun. Mater. DOI: 10.1002/adfm.201805822 (2019).

[23] L. J. Zhu, D. C. Ralph, and R. A. Buhrman, Irrelevance of magnetic proximity effect to the spin-orbit torques in heavy metal/ferromagnet bilayers, Phys. Rev. B **98**, 134406 (2018).

[24] W. Zhang, W. Han, X. Jiang, S.-H. Yang, S. S. P. Parkin, Role of transparency of platinum–ferromagnet interfaces in determining the intrinsic magnitude of the spin Hall effect, Nat. Phys. **11**, 496–502 (2015).

[25] L. Soumah, N. Beaulieu, L. Qassym, C. Carrétéro, E. Jacquet, R. Lebourgeois, J. B. Youssef, P. Bortolotti, V. Cros & A. Anane, Ultra-low damping insulating magnetic thin films get perpendicular, Nat. Commun. 9, 3355 (2018)

[26] M. C. Wheeler, F. Al MaMari, M. Rogers, F. J. Gonçalves, T. Moorsom, A. Brataas, R. Stamps, M. Ali, G. Burnell, B. J. Hickey & O. Cespedes, Optical conversion of pure spin currents in hybrid molecular devices, Nat. Comm. 8, 926 (2017).

[27] A. J. Berger, E. R. J. Edwards, H. T. Nembach, O. Karis, M. Weiler, and T. J. Silva, Determination of the spin Hall effect and the spin diffusion length of Pt from self-consistent fitting of damping enhancement and inverse spin-orbit torque measurements, Phys. Rev. B **98**, 024402 (2018).

[28] Y. Ou, S. Shi, D. C. Ralph, and R. A. Buhrman, Strong spin Hall effect in the antiferromagnet PtMn, Phys. Rev. B 93, 220405(R)(2016).

[29] R. Arias and D. L. Mills, Extrinsic contributions to the ferromagnetic resonance response of ultrathin films, Phys. Rev. B 60, 7395-7409 (1999).

[30] A. Azevedo, A. B. Oliveira, F. M. de Aguiar, and S. M. Rezende, Extrinsic contributions to spin-wave damping and renormalization in thin $Ni_{50}Fe_{50}$ films, Phys. Rev. B **62**, 5331-5333 (2000).

[31] Y. Liu, Z. Yuan, R. J. H. Wesselink, A. A. Starikov, and P. J. Kelly, Interface Enhancement of Gilbert Damping from First Principles, Phys. Rev. Lett. 113, 207202 (2014).

[32] K. Dolui and B. K. Nikolic, Spin-memory loss due to spin-orbit coupling at ferromagnet/heavy-metal interfaces: Ab initio spin-density matrix approach, Phys. Rev. B 96, 220403(R) (2017).

[33] L. Zhu, D. C. Ralph, R. A. Buhrman, Spin-Orbit Torques in Heavy-Metal–Ferromagnet Bilayers with Varying Strengths of Interfacial Spin-Orbit Coupling, Phys. Rev. Lett. 122, 077201 (2019).

[34] K. D. Belashchenko, Alexey A. Kovalev, and M. van Schilfgaarde, Theory of Spin Loss at Metallic Interfaces, Phys. Rev. Lett. 117, 207204 (2016).

[35] L. Zhu, D. C. Ralph, R. A. Buhrman, Enhancement of spin transparency by interfacial alloying, arXiv:1904.05455.

[36] L. Liu, T. Moriyama, D. C. Ralph, R. A. Buhrman, Spin-torque ferromagnetic resonance induced by the spin Hall effect, Phys. Rev. Lett. **106**, 036601 (2011).

[37] See Supplemental Materials for more information on sample preparation and measurement methods, determination of interfacial anisotropy energy density, numerical simulation of the magnetic damping due to two-magnon scattering, more examples on the FM thickness-dependence of damping of HM/FM bilayers, and magnetic roughness.

[38] M.-H. Nguyen, C.-F. Pai, K. X. Nguyen, D. A. Muller, D. C. Ralph, and R. A. Buhrman, Enhancement of the anti-damping spin torque efficacy of platinum by interface modification, Appl. Phys. Lett. 106, 222402 (2015).

[39] M. Zwierzycki, Y. Tserkovnyak, P. Kelly, A. Brataas, and G. E. W. Bauer, First-principles study of magnetization relaxation enhancement and spin transfer in thin magnetic films, Phys. Rev. B 71, 064420 (2005).

[40] C. O. Avci, K. Garello, M. Gabureac, A. Ghosh, A. Fuhrer, S. F. Alvarado, and P. Gambardella, Interplay of spin-orbit torque and thermoelectric effects in ferromagnet/ normal-metal bilayers, Phys. Rev. B **90**, 224427 (2014).

[41] H. L. Wang, C. H. Du, Y. Pu, R. Adur, P. C. Hammel, and F. Y. Yang, Scaling of Spin Hall Angle in 3d, 4d, and 5d Metals from $Y_3Fe_5O_{12}$/Metal Spin Pumping, Phys. Rev. Lett. 112, 197201 (2014).

[42] S. Ingvarsson, L. Ritchie, X. Y. Liu, Gang Xiao, J. C. Slonczewski, P. L. Trouilloud, and R. H. Koch, Role of electron scattering in the magnetization relaxation of thin $Ni_{81}Fe_{19}$ films, Phys. Rev. B 66, 214416 (2002).

[43] M. Oogane, T. Wakitani, S. Yakata, R. Yilgin, Y. Ando, A. Sakuma and T. Miyazaki, Magnetic Damping in Ferromagnetic Thin Films, Jpn. J. Appl. Phys. 45 3889 (2006).

[44] A. Conca, B. Heinz, M. R. Schweizer, S. Keller, E. Th. Papaioannou, and B. Hillebrands, Lack of correlation between the spin-mixing conductance and the inverse spin Hall effect generated voltages in CoFeB/Pt and CoFeB/Ta bilayers, Phys. Rev. B 95, 174426 (2007).

[45] T. K. H. Pham, M. Ribeiro, J. H. Park, N. J. Lee, K. Hoon Kang, E. Park, V. Q. Nguyen, A. Michel, C. S. Yoon, S. Cho, and T. H. Kim, Interface morphology effect on the spin mixing conductance of Pt/$Fe_3O_4$ bilayers, Sci. Rep. **8**, 13907 (2018).

[46] X. Ma, G. Yu, C. Tang, X. Li, C. He, J. Shi, K. L. Wang, and X. Li, Interfacial Dzyaloshinskii-Moriya Interaction: Effect of 5*d* Band Filling and Correlation with Spin Mixing Conductance, Phys. Rev. Lett. 120, 157204 (2018).

[47] C.-F. Pai, M.-H. Nguyen, C. Belvin, L. H. Vilela-Leão, D. C. Ralph, and R. A. Buhrman, Enhancement of perpendicular magnetic anisotropy and transmission of spin-Hall-effect-induced spin currents by a Hf spacer layer in W/Hf/CoFeB/MgO layer structures, Appl. Phys. Lett. 104, 082407 (2014).